\documentclass[10pt]{iopart}

\usepackage{graphicx}
\usepackage{iopams}

\begin{document}

\title{Absence of superconductivity in the two-dimensional Hubbard model}

\author{A Sherman}

\address{Institute of Physics, University of Tartu, W. Ostwaldi Str 1, 50411 Tartu, Estonia}

\ead{alekseis@ut.ee}

\begin{abstract}
The possibility of the superconducting transition in the two-dimensional repulsive Hubbard model is studied using the ladder approximation of the strong coupling diagram technique. The $t$-$U$ and $t$-$t'$-$t''$-$U$ models are considered in the regime of strong correlations, for the on-site Coulomb repulsion $U=8t$, in the range of temperatures $0.02t\lesssim T\lesssim 0.3t$. To avoid the influence of the phase separation and size effects the calculations are performed in an infinite crystal, in the part of the phase diagram without inhomogeneities, for the electron concentration $\bar{n} =0.92$. Solutions of the Eliashberg equation for singlet and triplet pairing, which are transformed according to one-dimensional representations of the lattice point group $D_4$, are considered. For both models and all considered symmetries, eigenvalues of the Eliashberg equation are less than unity and demonstrate no upward trend with decreasing temperature. This result points to the absence of superconductivity in the Hubbard model in the strong-coupling regime. We discuss the reason for the small eigenvalue of the Eliashberg equation in the case of the singlet $d_{x^2-y^2}$ pairing.
\end{abstract}

\vspace{2pc}


\maketitle

\ioptwocol

\section{Introduction}
From the early stages of the study of high-temperature superconductivity, the single-band two-dimensional (2D) Hubbard model was argued to be the paradigmatic model for the problem \cite{Anderson,Zhang}. Indeed, many properties of cuprate perovskites are correctly described in the model. The most prominent of them are antiferromagnetism and its doping dependence, charge instability, and pseudogap formation (see, e.g., \cite{Hirsch,Scalapino,Otsuki,Aichhorn,Macridin,White,Sherman18,Sherman19a,Sherman20,Bill}). As for the ability to describe superconductivity, results are more diverse -- both affirmative and negative answers were obtained depending on used methods (see, e.g., \cite{Scalapino,Otsuki,Senechal,Capone,Aimi,Maier,Kitatani,Qin}). The absence of superconductivity is usually related to its competition with the stripe formation. In this connection, the known experimental fact should be mentioned: fluctuating stripes coexist with superconductivity; it is suppressed by static stripes \cite{Tranquada,Forgan}. Since sample boundaries can pin stripes, the use of comparatively small clusters can influence the answer. There is one more problem with cluster approaches. As will be seen below, the spin vertex is an essential part of the matrix in the Eliashberg equation \cite{Eliashberg}. The behavior of this vertex at low frequencies near the antiferromagnetic ordering vector plays the central role for the eigenvalue and symmetry of the solution. The antiferromagnetic order is established in a cluster as the magnetic correlation length approaches the cluster size. As a consequence, for moderate temperatures, the vertex in the cluster is larger than in an infinite crystal.

In this work, we study solutions of the Eliashberg equation derived for the Hubbard model in the framework of the strong coupling diagram technique (SCDT) \cite{Vladimir,Metzner,Pairault,Sherman16}. This approach is the regular series expansion in powers of hopping constants around the atomic limit. The approach can describe the Mott metal-insulator transition \cite{Zaitsev,Izyumov,Vladimir,Pairault} and, therefore, in contrast to the usual weak coupling diagram technique, it is applicable for the case of strong electron correlations. For the 2D one-band Hubbard model, the validity of the SCDT was controlled in comparison with the results of numeric experiments and experiments with ultracold fermionic atoms in 2D optical lattices. In particular, it was shown that the critical repulsion for the Mott metal-insulator transition is close to that observed in Monte-Carlo simulations. For the comparable parameters, spectral functions and den\-si\-ties of states are similar to those found in ex\-act diagonalizations and Monte Carlo simulations \cite{Sherman18}. Temperature and concentration dependencies of the uniform spin susceptibility, spin structure fac\-tor, square of the site spin, and double occupancy are in good agreement with re\-sults of Monte Carlo simulations, numeric linked-cluster expansion, and experiments with ul\-tra\-cold fermionic atoms \cite{Sherman18,Sherman19a}. Shapes and intensity distributions in Fermi surfaces in electron- and hole-doped cases are similar to those observed experimentally \cite{Sherman18,Sherman19b}. Lastly, moments sum rules are fulfilled with good accuracy \cite{Sherman18}.

Calculations in the SCDT are not more complicated than those in the weak coupling diagram technique. In particular, the linked-cluster theorem is valid, and partial summations are allowed in SCDT. For calculating the superconducting susceptibility, it appears reasonable to approximate particle-particle irreducible diagrams by infinite sums of ladder diagrams describing spin and charge fluctuations. These sums form spin and charge vertices defining spin and charge susceptibilities. In the present work, the ladder diagrams are constructed from second-order cumulants of electron operators and renormalized hopping lines containing electron Green's functions. These functions are self-consistently calculated using the same approach as in our earlier works \cite{Sherman18,Sherman19a,Sherman20,Bill,Sherman19b}. We consider both the $t$-$U$ model, with the non-zero hopping constant $t$ between nearest-neighbor sites only, and the $t$-$t'$-$t''$-$U$ model, with additional non-zero hopping constants between second and third neighbors. The Hubbard model exhibits phase separation when the chemical potential $\mu$ is near zero \cite{Sherman20,Bill}. For the considered regime of strong correlations with the on-site Coulomb repulsion $U=8t$, this instability occurs near electron concentrations $\bar{n}=0.8$ in the $t$-$U$ model and near $\bar{n}=0.87$ in the $t$-$t'$-$t''$-$U$ model with $t'=-0.3t$, $t''=0.2t$, and the temperature $T\approx 0.1t$. These values of $t'$ and $t''$ were suggested by band-structure calculations \cite{Andersen}. To avoid the influence of this instability on the solution, the chemical potential is chosen to lie far from zero such that $\bar{n}\approx 0.92$. To get rid of the effect of cluster boundaries, calculations are carried out for an infinite crystal. Order parameters corresponding to singlet and triplet pairing of all one-dimensional representations of the $D_4$ point group of the system are considered.

For both considered models, in the range of temperatures $0.02t\lesssim T\lesssim 0.3t$, eigenvalues of the Eliashberg equation appear to be smaller than unity and demonstrate no tendency to grow with decreasing temperature. This result points to the absence of superconductivity. For the singlet $B_1$ ($d_{x^2-y^2}$) even-frequency pairing, which is of particular interest, the value of the eigenvalue is determined by maxima of the spin vertex at low frequencies and momenta near the corner of the Brillouin zone. In contrast to half-filling, with decreasing $T$, the intensity of these maxima is modest and grows only moderately for finite doping, which explains a small value of the eigenvalue. The same components of the spin vertex determine the zero-frequency staggered spin susceptibility. Its modest increase and saturation with decreasing $T$ explain the known experimental fact that for finite doping and low temperatures, the magnetic correlation length ceases to depend on $T$ and is determined by the carrier concentration \cite{Keimer}. Using the value of the superexchange constant $J=4t^2/U=0.1$~eV, as observed in cuprates, for $U=8t$, we find $t=0.2\,{\rm eV}\approx 2000$~K. Thus, the lowest considered temperature $0.02t\approx 40$~K, which is close to the superconducting transition temperature in La$_{2-x}$Ba$_x$CuO$_4$. We conclude that the Hubbard model cannot explain transition temperatures in this range.

The paper is organized as follows. In Section~2, main equations are derived, and a brief consideration of their solution is given. The main results are discussed in Section~3. The last section is devoted to concluding remarks.

\section{Model and SCDT method}
The Hamiltonian of the 2D fermionic Hubbard model \cite{Hubbard63,Hubbard64} reads
\begin{equation}\label{Hamiltonian}
H=\sum_{\bf ll'\sigma}t_{\bf ll'}a^\dagger_{\bf l'\sigma}a_{\bf l\sigma}
+\frac{U}{2}\sum_{\bf l\sigma}n_{\bf l\sigma}n_{\bf l,-\sigma},
\end{equation}
where 2D vectors ${\bf l}$ and ${\bf l'}$ label sites of a square plane lattice, $\sigma=\uparrow,\downarrow$ is the spin projection, $a^\dagger_{\bf l\sigma}$ and $a_{\bf l\sigma}$ are electron creation and annihilation operators, $t_{\bf ll'}$ is the hopping constant and $n_{\bf l\sigma}=a^\dagger_{\bf l\sigma}a_{\bf l\sigma}$ is the number operator. As mentioned in the Introduction, in this work, two cases of hopping constants are considered. In one of them, only the integral between nearest-neighbor sites $t$ is nonvanishing. In the second case, the integrals between second $t'=-0.3t$ and third $t''=0.2t$ neighbors are also taken into account \cite{Andersen}.

For investigating the superconductivity, the zero-frequency homogeneous superconducting susceptibility \cite{Scalapino},
\begin{eqnarray}
\chi^{\rm sc}&=&\frac{1}{N}\int_{0}^{\beta}\langle{\cal T}\Delta(\tau)\Delta^\dagger\rangle{\rm d}\tau \nonumber\\
&=&\frac{1}{N}\int_{0}^{\beta}\sum_{\bf lm}\sum_{\bf l'm'}\phi_{\bf lm}\phi^*_{\bf l'm'}\label{suscept}\\
&&\quad\times\chi({\bf m}\downarrow\tau,{\bf l\uparrow\tau},{\bf l'}\uparrow0,{\bf m'}\downarrow0)\,{\rm d}\tau,\nonumber
\end{eqnarray}
will be considered. Here
\begin{eqnarray}\label{chi}
&&\chi({\bf m}\downarrow\xi,{\bf l}\uparrow\tau,{\bf l'}\uparrow\tau',{\bf m'\downarrow\xi'})\nonumber\\
&&\quad\quad=\langle{\cal T} a_{\bf m\downarrow}(\xi)a_{\bf l\uparrow}(\tau)a^\dagger_{\bf l'\uparrow}(\tau')a^\dagger_{\bf m'\downarrow}(\xi')\rangle,
\end{eqnarray}
$\Delta=\sum_{\bf lm}\phi_{\bf lm}a_{\bf m\downarrow}a_{\bf l\uparrow}$, $\phi_{\bf lm}$ is the pairing function, $\beta=1/T$ is the inverse tem\-pe\-ra\-tu\-re, ${\cal T}$ is the chronological operator, the statistical averaging denoted by angle brackets and operator time dependencies are determined by the Hamiltonian ${\cal H}=H-\mu\sum_{\bf l\sigma}a^\dagger_{\bf l\sigma}a_{\bf l\sigma}$, $N$ is the number of sites.

\begin{figure}[t]
\centerline{\resizebox{0.99\columnwidth}{!}{\includegraphics{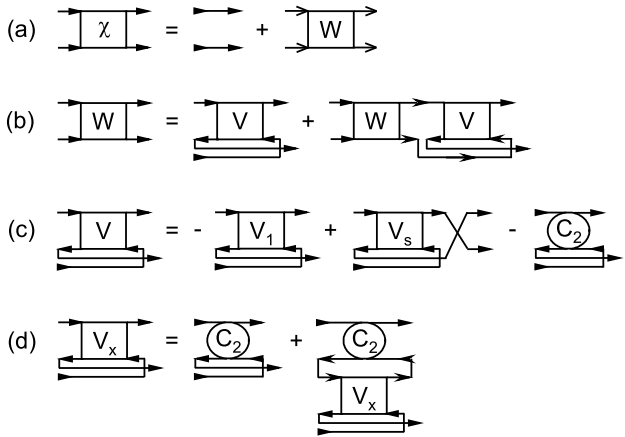}}}
\caption{The diagrammatic representation of main equations. In parts (a)--(c), the upper legs of the four-leg diagrams are characterized by the spin projections $\sigma=\uparrow$, while the lower ones by  $\sigma=\downarrow$. In part (d), $V_x$ is either $V^s$ or $V^c$ and $C_2$ is the respective second-order cumulant.} \label{Fig1}
\end{figure}

In the SCDT \cite{Vladimir,Metzner,Pairault,Sherman16}, many-particle correlation functions of the type of quantity~(\ref{chi}) are calculated from the series expansion in powers of hopping constants of the kinetic term of the Hamiltonian. The approach is well suited for the strong electron correlations and is visualized by the diagram technique. The terms of the expansion are products of hopping constants and cumulants \cite{Kubo} of electron operators, which are depicted by directed lines and circles, respectively. The power expansion of the susceptibility (\ref{chi}) contains two groups of diagrams. In one of these groups, cumulants and hopping lines connect ${\bf l'}\tau'$ and ${\bf l}\tau$ endpoints of the correlator, while other cumulants and lines join ${\bf m'}\xi'$ and ${\bf m}\xi$ points, without any elements connecting these two diagram sequences. The sum of all such diagrams gives the product of two electron Green's functions with respective endpoints. In the other group of diagrams, all four ends of $\chi$ are in some manner connected. In such a diagram, endpoints can be connected to the rest of the diagram by any sequence of hopping lines and cumulants. The sum of all such sequences produces the terminal line $\Pi$. We denote the sum of all diagrams of the second group without the four terminal lines as $W$. As a consequence, equations for susceptibilities (\ref{suscept}) and (\ref{chi}) read
\begin{eqnarray}\label{ftchi}
&&\chi(p_2,\downarrow;p_1+p_3-p_2,\uparrow;p_1,\uparrow;p_3,\downarrow)\nonumber\\ &&\quad=\frac{N}{T}\delta_{p_3p_2}G(p_1)G(p_3)\nonumber\\
&&\quad\quad+\Pi(p_1)\Pi(p_2)\Pi(p_3)\Pi(p_1+p_3-p_2)\nonumber\\
&&\quad\quad\times W(p_2,\downarrow;p_1+p_3-p_2,\uparrow;p_1,\uparrow;p_3,\downarrow),
\end{eqnarray}
\begin{eqnarray}\label{ftsuscept}
\chi^{\rm sc}&=&\frac{T}{N}\sum_p|\phi_{\bf k}|^2G(p)G(-p)\nonumber \\
&&+\bigg(\frac{T}{N}\bigg)^2\sum_{pp'}\phi^*_{\bf k'}\phi_{\bf k}\Pi(p)\Pi(-p)\Pi(p')\Pi(-p')\nonumber \\
&&\quad\times W(-p,\downarrow;p,\uparrow;p',\uparrow,-p'\downarrow),
\end{eqnarray}
where the Fourier transformation was employed, $p$ is the variable containing the 2D wave vector {\bf k} and Matsubara frequency $\omega_j=(2j-1)\pi T$ with an integer $j$, $G(p)$ is the electron Green's function, and the terminal line $\Pi(p)=1+t_{\bf k}G(p)$ with the Fourier transform of the hopping constants $t_{\bf k}$. The graphic representation of equation (\ref{ftchi}) is shown in figure~\ref{Fig1}(a). Here arrows in the vertices of the four-leg diagrams denote their input and output ends. Lines with two arrows at ends are electron Green's functions. To distinguish terminal lines $\Pi$ connected to the four-leg diagram $W$ from input-output ends, we use arrows with different heads.

In its turn, diagrams forming $W$ can be separated into two-particle reducible and irreducible ones. The latter cannot be divided into two disconnected parts by cutting two horizontal hopping lines pointed in the same direction. If we denote the sum of all two-particle irreducible diagrams as $V$, the vertex $W$ can be described by the following Bethe-Salpeter equation (BSE):
\begin{eqnarray}\label{W}
&&W(-p,\downarrow;p,\uparrow;p',\uparrow;-p',\downarrow)\nonumber\\
&&\quad=V(p',\uparrow;-p,\downarrow;-p',\downarrow;p,\uparrow) \nonumber \\
&&\quad\quad+\frac{T}{N}\sum_{p''}
V(p',\uparrow;-p'',\downarrow;-p',\downarrow;p'',\uparrow) \nonumber \\
&&\quad\quad\times\theta(p'')\theta(-p'')
W(-p,\downarrow;p,\uparrow;p'',\uparrow;-p'',\downarrow).
\end{eqnarray}
The enumeration of legs in $V$ differs from that in $W$. It is to make a straightforward connection with spin and charge vertices of the previous SCDT calculations \cite{Sherman18,Sherman19a}. In this approach, a two-leg diagram is termed one-particle irreducible if it cannot be divided into two disconnected parts by cutting a hopping line. Due to the possibility of the partial summation in SCDT, all possible sequences of one-particle irreducible diagrams can be inserted in the internal hopping lines in $W$. As a result, the bare hopping lines $t_{\bf k}$ are transformed to the renormalized hopping $\theta(p)=t_{\bf k}+t_{\bf k}^2G(p)$. These quantities enter into (\ref{W}). The graphical representation of this equation is shown in figure~\ref{Fig1}(b), in which the renormalized hopping lines connect $V$ and $W$ four-leg vertices (other arrowheads distinguish $\theta$ from vertex endpoints and $\Pi$).

In this work, we approximate the two-particle irreducible vertex $V$ by infinite sums of ladder diagrams. In the Hubbard model, there are two types of such sums, which differ in spin indices of their endpoints. We denote them $V^s$ and $V^c$. These quantities describe spin and charge fluctuations and determine spin and charge susceptibilities. In the present work, the ladder diagrams are constructed from second-order cumulants $C^{(2)}$ of electron operators and renormalized hopping lines $\theta$. It was shown \cite{Sherman18,Sherman19a} that electron spectral functions and susceptibilities calculated in this approximation agree well with Monte Carlo simulations and experiments with ultracold atoms. The sum $V^s$ satisfies the following BSE \cite{Sherman18,Sherman19a}:
\begin{eqnarray}
&&V^s_{\bf k}(j+\nu,j,j',j'+\nu)=C^{(2a)}(j+\nu,j,j',j'+\nu)\nonumber \\
&&\quad+T\sum_{\nu'}C^{(2a)}(j+\nu,j+\nu',j'+\nu',j'+\nu) \nonumber\\
&&\quad\quad\times {\cal T}_{\bf k}(j+\nu',j'+\nu')V^s_{\bf k}(j+\nu',j,j',j'+\nu'),\label{Vs}
\end{eqnarray}
where ${\cal T}_{\bf k}(j,j')=N^{-1}\sum_{\bf k'}\theta({\bf k'+k},j)\theta({\bf k'},j')$, the shorthand notation $j$ stands for the frequency $\omega_j$, $\nu$ is an integer, and $C^{(2a)}$ is the second-order cumulant antisymmetrized over spin indices,
\begin{eqnarray*}
&&C^{(2a)}(j+\nu,j,j',j'+\nu)\\
&&\quad=\sum_{\sigma'}\sigma\sigma'
C^{(2)}(j+\nu,\sigma';j,\sigma;j,\sigma;j'+\nu,\sigma').
\end{eqnarray*}
The sum of ladder diagrams $V^c$ are described by the BSE similar to (\ref{Vs}) with $C^{(2a)}$ substituted by the cumulant $C^{(2s)}$ symmetrized over spin indices,
\begin{eqnarray*}
&&C^{(2s)}(j+\nu,j,j',j'+\nu)\\
&&\quad=\sum_{\sigma'}C^{(2)}(j+\nu,\sigma';j,\sigma;j,\sigma;j'+\nu,\sigma').
\end{eqnarray*}
These BSEs are depicted in figure~\ref{Fig1}(d), where $V_x$ is either $V^s$ or $V^c$ and $C_2$ is the second-order cumulant (anti)symmetrized over spin indices.

In the above equations, the second-order cumulant $C^{(2)}$ reads
\begin{eqnarray*}
&&C^{(2)}(\tau_1,\sigma';\tau_2,\sigma;\tau_3,\sigma;\tau_4,
\sigma')\\
&&\quad=\big\langle{\cal T}a^\dagger_{\sigma'}(\tau_1)a_{\sigma}(\tau_2) a^\dagger_{\sigma}(\tau_3)a_{\sigma'}(\tau_4)\big\rangle_0\\
&&\quad\quad -\big\langle{\cal T}a^\dagger_{\sigma'}(\tau_1)a_{\sigma}(\tau_2)\big\rangle_0 \big\langle{\cal T}a^\dagger_{\sigma}(\tau_3)a_{\sigma'}(\tau_4)\big\rangle_0 \delta_{\sigma\sigma'}\\
&&\quad\quad+\big\langle{\cal T}a^\dagger_{\sigma'}(\tau_1)a_{\sigma'}(\tau_4)\big\rangle_0 \big\langle{\cal T}a^\dagger_{\sigma}(\tau_3)a_{\sigma}(\tau_2)\big\rangle_0,
\end{eqnarray*}
where the subscript 0 near angle brackets indicates that time dependencies and the thermodynamic averaging are determined by the Hamiltonian of the Hubbard atom
\begin{equation}\label{site}
H_{\bf l}=\sum_\sigma\bigg(\frac{U}{2}n_{\bf l\sigma}n_{\bf l,-\sigma}-\mu n_{\bf l\sigma}\bigg).
\end{equation}
Due to the translation symmetry, the cumulant does not depend on the site index, which is, therefore, omitted. The Fourier transform of the cumulant reads \cite{Vladimir,Sherman18}
\begin{eqnarray}\label{C2}
&&C^{(2)}(j+\nu,\sigma';j,\sigma;j',\sigma;j'+\nu,\sigma')\nonumber\\
&&\quad=-\Big[Z^{-1}{\rm e}^{-\beta E_1}\beta(\delta_{jj'}\delta_{\sigma\sigma'}- \delta_{\nu0})+Z^{-2}\Big({\rm e}^{-\beta(E_0+E_2)}\nonumber\\
&&\quad-{\rm e}^{-2\beta E_1}\Big)\beta(\delta_{jj'} -\delta_{\nu0}\delta_{\sigma\sigma'})\Big] F(j)F(j'+\nu)\nonumber\\
&&\quad+Z^{-1}{\rm e}^{-\beta E_0}\delta_{\sigma,-\sigma'}Ug_{01}(j)g_{01}(j'+\nu) \nonumber\\
&&\quad\times g_{02}(\omega_j+\omega_{j'+\nu})[g_{01}(j')+g_{01}(j+\nu)]\nonumber\\
&&\quad+Z^{-1}{\rm e}^{-\beta E_2}\delta_{\sigma,-\sigma'}Ug_{12}(j)g_{12}(j'+\nu) \nonumber\\
&&\quad\times g_{02}(\omega_j+\omega_{j'+\nu})[g_{12}(j')+g_{12}(j+\nu)]\nonumber\\
&&\quad-Z^{-1}{\rm e}^{-\beta E_1}\delta_{\sigma,-\sigma}\{F(j'+\nu)[g_{01}(j)g_{01}(j') \nonumber\\
&&\quad+g_{12}(j)g_{12}(j+\nu)-g_{01}(j')g_{12}(j+\nu)]\nonumber\\
&&\quad+F(j)[g_{01}(j'+\nu)g_{01}(j+\nu)+g_{12}(j'+\nu)g_{12}(j') \nonumber\\
&&\quad-g_{01}(j+\nu)g_{12}(j')]\},
\end{eqnarray}
where $E_0=0$, $E_1=-\mu$, and $E_2=U-2\mu$ are energies of the empty, singly, and doubly occupied states of the Hamiltonian (\ref{site}), $g_{ii'}(j)=g_{ii'}(\omega_j)=({\rm i}\omega_j+E_i-E_{i'})^{-1}$, the atomic partition function $Z=\exp(-\beta E_0)+2\exp(-\beta E_1)+\exp(-\beta E_2)$, and $F(j)=g_{01}(j)- g_{12}(j)$.

Thanks to the Boltzmann factors $\exp(-\beta E_i)$ in (\ref{C2}), the cumulant is substantially simplified in the range of chemical potentials $\mu\gg T$, $U-\mu\gg T$. This range contains cases of half-filling and moderate doping. The regions of the charge instability \cite{Sherman20}, which influence we wish to exclude from consideration, do not fall into this range. This cumulant simplification reduces BSE (\ref{Vs}) to four linear equations with four unknowns, parametrically dependent on frequencies and the momentum \cite{Sherman18}. This equation set is easily solved. The second BSE for $V^c$ is solved similarly.

The two-particle irreducible vertex $V$ is expressed through the spin and charge vertices as
\begin{eqnarray}\label{V}
&&V(p',\uparrow;-p,\downarrow;-p',\downarrow;p,\uparrow)\nonumber\\
&&\quad=-V^1_{\bf k'-k}(\omega_{j'},\uparrow;-\omega_j,\downarrow;-\omega_{j'}, \downarrow;\omega_j,\uparrow)\nonumber \\
&&\quad+V^s_{\bf -k'-k}(-\omega_{j'},\downarrow;-\omega_j,\downarrow;\omega_{j'}, \uparrow;\omega_j,\uparrow)\nonumber\\
&&\quad-C^{(2a)}(-\omega_{j'},-\omega_j,\omega_{j'},\omega_j),
\end{eqnarray}
where
\begin{eqnarray*}
&&V^1_{\bf k}(\omega_{j'},\uparrow;-\omega_j,\downarrow;-\omega_{j'}, \downarrow;\omega_j,\uparrow) \\
&&\quad=\frac{1}{2}\big[V^c_{\bf k}(\omega_{j'},-\omega_j,-\omega_{j'}, \omega_j)\\
&&\quad-V^c_{\bf k}(\omega_{j'},-\omega_j,-\omega_{j'}, \omega_j)\big].
\end{eqnarray*}
The last term on the right-hand side of (\ref{V}) compensates double counting of the same diagram. Graphically this equation is shown in figure~\ref{Fig1}(c).

It is convenient to introduce two new quantities,
\begin{eqnarray*}
W^{\pm}_{p'p}&=&W(-p,\downarrow;p,\uparrow;p',\uparrow;-p',\downarrow)\\
&&\pm W(-p,\downarrow;p,\uparrow;-p',\uparrow;p',\downarrow),\\
V^{\pm}_{p'p}&=&V(p',\uparrow;-p,\downarrow;-p',\downarrow;p,\uparrow)\\
&&\pm V(-p',\uparrow;-p,\downarrow;p',\downarrow;p,\uparrow).
\end{eqnarray*}
Quantities with the plus sign correspond to singlet pairing, while those with the minus sign are connected with triplet pairing. As follows from the above equations, $V^{\pm}$ possesses the following properties:
\begin{equation*}
V^{\pm}_{-p',p}=\pm V^{\pm}_{p'p}=V^{\pm}_{p',-p}=\pm V^{\pm}_{-p',-p}.
\end{equation*}
Using these properties, equation (\ref{W}) can be rewritten as
\begin{equation}\label{Wpm}
W^{\pm}_{p'p}=V^{\pm}_{p'p}+\frac{T}{2N}\sum_{p''}V^{\pm}_{p'p''}\theta_{p''} \theta_{-p''}W^{\pm}_{p''p}
\end{equation}
with
\begin{eqnarray}
V^+_{p'p}&=&-\frac{1}{2}\big[V^c_{\bf k'-k}(\omega_{j'},-\omega_j,-\omega_{j'}, \omega_j)\nonumber\\
&&\quad+V^c_{\bf -k'-k}(-\omega_{j'},-\omega_j,\omega_{j'},\omega_j)\big] \nonumber\\
&&+\frac{3}{2}\big[V^s_{\bf k'-k}(\omega_{j'},-\omega_j,-\omega_{j'}, \omega_j)\nonumber\\
&&\quad+V^s_{\bf -k'-k}(-\omega_{j'},-\omega_j,\omega_{j'},\omega_j)\big] \nonumber\\
&&-C^{2a}(\omega_{j'},-\omega_j,-\omega_{j'}, \omega_j)\nonumber\\
&&-C^{2a}(-\omega_{j'},-\omega_j,\omega_{j'},\omega_j),\nonumber\\[-1ex]
&&\label{Vpm}\\[-1ex]
V^-_{p'p}&=&\frac{1}{2}\big[-V^c_{\bf k'-k}(\omega_{j'},-\omega_j,-\omega_{j'}, \omega_j)\nonumber\\
&&\quad+V^c_{\bf -k'-k}(-\omega_{j'},-\omega_j,\omega_{j'},\omega_j)\nonumber\\
&&\quad-V^s_{\bf k'-k}(\omega_{j'},-\omega_j,-\omega_{j'}, \omega_j)\nonumber\\
&&\quad+V^s_{\bf -k'-k}(-\omega_{j'},-\omega_j,\omega_{j'},\omega_j)\big] \nonumber\\
&&+C^{2a}(\omega_{j'},-\omega_j,-\omega_{j'}, \omega_j)\nonumber\\
&&-C^{2a}(-\omega_{j'},-\omega_j,\omega_{j'},\omega_j).\nonumber
\end{eqnarray}
From the above equations, it follows that the matrices $V^{\pm}_{p'p}\theta_p\theta_{-p}$ are real, which, however, does not assure that their eigenvalues are real since the matrices are nonsymmetric.

The Eliashberg equation corresponding to (\ref{Wpm}) reads
\begin{equation}\label{Eliash}
\frac{T}{2N}\sum_{p}V^{\pm}_{p'p}\theta_p\theta_{-p}\varphi_{p}=E\varphi_{p'}.
\end{equation}
The quantity $W^{\pm}$ and, along with it, the susceptibility (\ref{suscept}) diverges when one of the eigenvalues of the matrix equation (\ref{Eliash}) becomes equal to unity. It is the condition for the superconducting transition.

Momentum dependencies of the matrices in (\ref{Eliash}) are invariant with respect to the $D_4$ point group of the lattice. This group has five representations, four of which -- $A_1$ ($x^2+y^2$), $A_2$ ($z$), $B_1$ ($x^2-y^2$), and $B_2$ ($xy$) -- are one-dimensional and one is two-dimensional \cite{Bir}. Under point group transformations, the functions $\varphi_{p}$ are changed in accordance with one of these representations. We limit ourselves to the one-dimensional representations.

To solve the Eliashberg equation (\ref{Eliash}) we used the power (von Mises) iteration \cite{Mises}. Electron Green's functions, which were necessary for calculating the matrix in this equation, were found by the same method as used in earlier works \cite{Sherman18,Sherman19a,Sherman19b}. The above equations, as well as equations of these works, were derived for an infinite crystal. The discretization of the momentum space is used for an approximate integration over {\bf k} only. It allows us to avoid the influence of size effects on the superconducting transition.

\section{Results and discussion}
\begin{figure}[t]
\centerline{\resizebox{0.99\columnwidth}{!}{\includegraphics{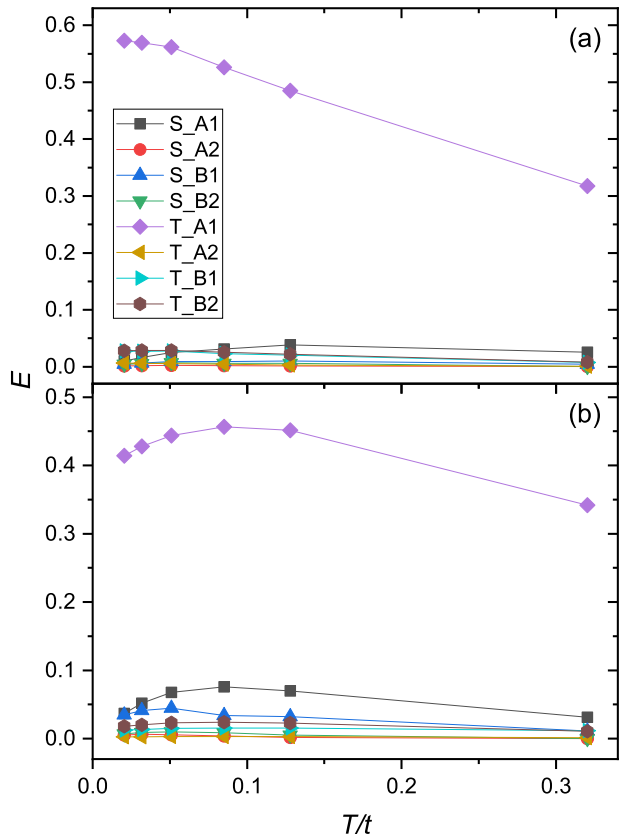}}}
\caption{Eigenvalues of Eliashberg equation (\protect\ref{Eliash}) as functions of temperature for $U=8t$, $t_1=t_2=0$ (a) and $t_1=-0.3t$, $t_2=0.2t$ (b). For both models $\bar{n}=0.92$. Lines and symbols of different colors correspond to different pairing symmetries indicated in the legend, in which letter S points to singlet and T to triplet pairing.} \label{Fig2}
\end{figure}
Eigenvalues of the Eliashberg equation (\ref{Eliash}) as functions of temperature are shown in figure~\ref{Fig2} for the $t$-$U$ and $t$-$t'$-$t''$-$U$ models. They are much smaller than unity and do not increase rapidly with decreasing temperature. The largest eigenvalue corresponds to the exotic odd-frequency triplet $A_1$ pairing. It results from charge fluctuations -- the eigenvalue decreases sharply if $V^c$ is dropped in $V^-$ (\ref{Vpm}). Thus, the superconducting transition does not occur for the considered symmetries.

\begin{figure}[t]
\centerline{\resizebox{0.99\columnwidth}{!}{\includegraphics{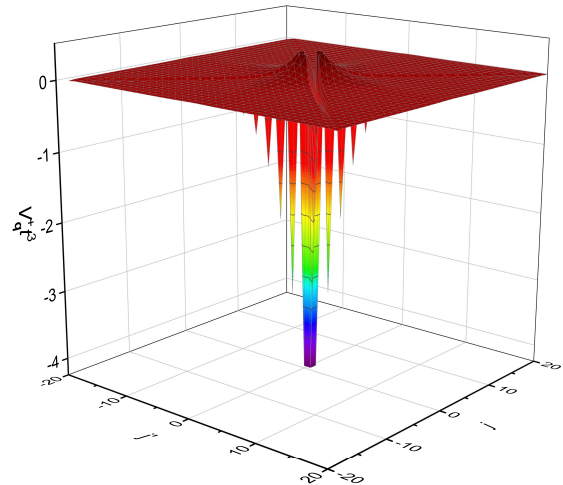}}}
\caption{The frequency dependence of $V^+_{\bf q}(j',j)$ for $U=8t$, $t'=-0.3t$, $t''=0.2t$, $T\approx 0.13t$, $\bar{n}=0.92$, and ${\bf q}=(\pi,\pi)$.} \label{Fig3}
\end{figure}

What is the mechanism, which suppresses superconductivity in the Hubbard model? We chose the part of the phase diagram without charge instabilities. Thus, this reason can be excluded from consideration. Of primary interest is the singlet $d_{x^2-y^2}$ ($B_1$) pairing and the contribution of spin fluctuations to it. Notice that the matrix for singlet pairing $M_{p'p}=(T/2N)V_{p'p}^+\theta_p\theta_{-p}$ in the Eliashberg equation (\ref{Eliash}) is sharply peaked at small frequencies for all momenta. Figure~\ref{Fig3} demonstrates this matrix property.
In this figure, we take into account that for one-dimensional representations of the $D_4$ group vertices $V^s_{\bf -k'-k}$ and $V^c_{\bf -k'-k}$ in $V^\pm_{p'p}$ (\ref{Vpm}) can be substituted with $V^s_{\bf k'-k}$ and $V^c_{\bf k'-k}$, respectively. Hence $V^\pm_{p'p}$ depends only on the following three variables: ${\bf q=k'-k}$, $j'$, and $j$. In the figure caption, the value of the wave vector is given in units of the inverse lattice spacing. Thus, the low-frequency region of the matrix $M$ determines the eigenvalue of the equation (\ref{Eliash}).

\begin{figure}[tbh]
\centerline{\resizebox{0.99\columnwidth}{!}{\includegraphics{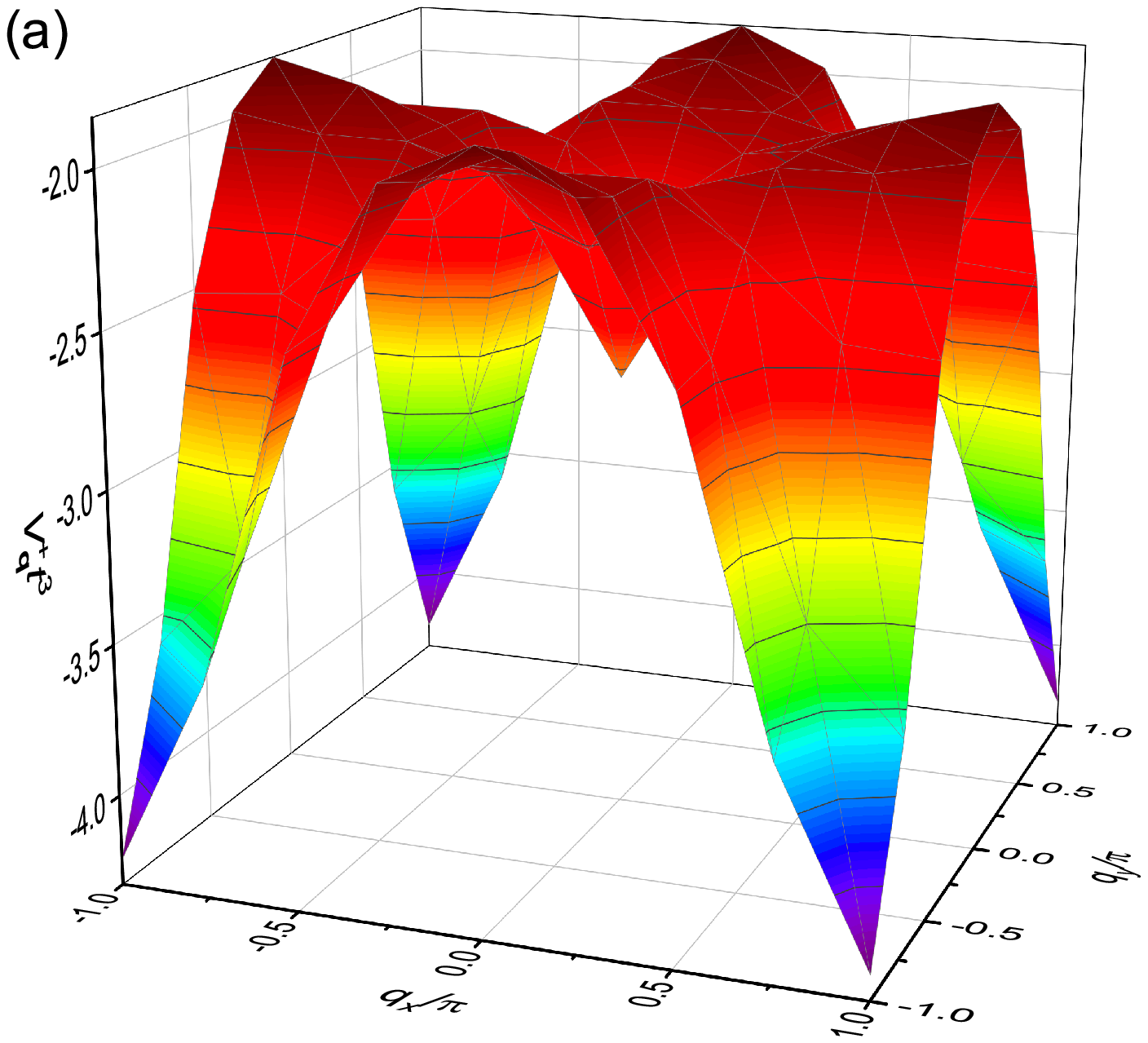}}}
\vspace{1ex}
\centerline{\resizebox{0.99\columnwidth}{!}{\includegraphics{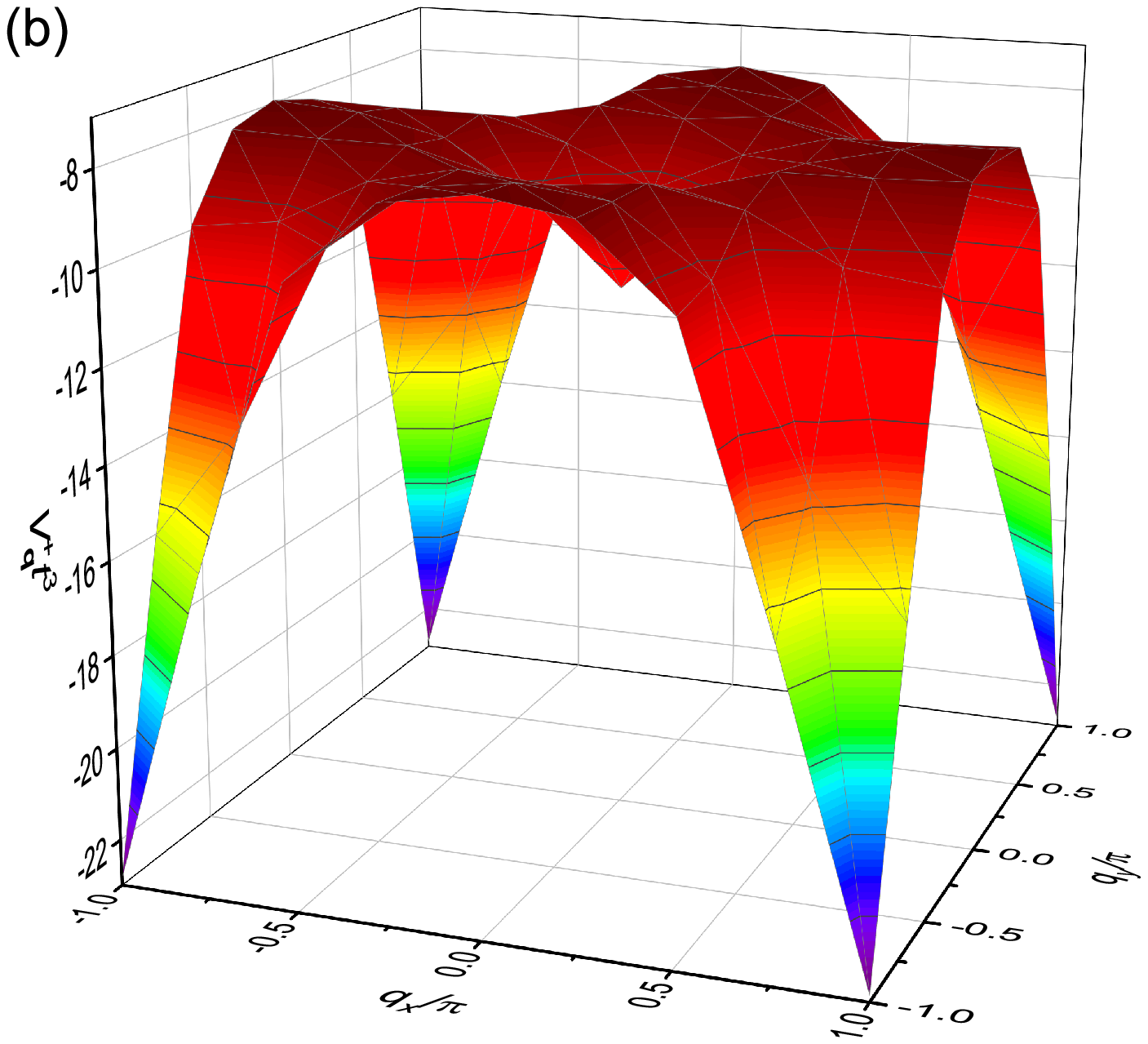}}}
\caption{The momentum dependence of $V^+_{\bf q}(j',j)$ for $U=8t$, $t'=-0.3t$, $t''=0.2t$, $\bar{n}=0.92$, and $j'=j=0$. Temperatures for panels (a) and (b) are $T\approx 0.13t$ and $0.032t$, respectively.} \label{Fig4}
\end{figure}
The momentum dependence of the vertex $V^+_{\bf q}(j',j)$ is shown in figure~\ref{Fig4} for two values of temperature. The dependence has the pronounced minimum at the momentum of the antiferromagnetic ordering ${\bf q=Q}=(\pi,\pi)$, which is connected with the spin vertex $V^s$ in (\ref{Vpm}). A less pronounced local minimum at ${\bf q}=(0,0)$ is caused by the charge vertex $V^c$. At a large enough difference $V^+_{\bf 0}-V^+_{\bf Q}$, such shape of the vertex promotes a large eigenvalue
\begin{equation*}\label{eigen}
E=\frac{\sum_{{\bf k'}j'}\sum_{{\bf k}j}\varphi^*_{\bf k'}(j')M_{\bf k'-k}(j',j)\varphi_{\bf k}(j)}{\sum_{{\bf k}j}\varphi^*_{\bf k}(j)\varphi_{\bf k}(j)}
\end{equation*}
for the $d_{x^2-y^2}$ pairing. Indeed, in this case, the maximum at ${\bf k'}=(\pi,0)$ in the function $\varphi^*_{\bf k'}(j')$ and the minimum at ${\bf k}=(0,-\pi)$ in the function $\varphi_{\bf k}(j)$ fall on the minimum of $M_{\bf k'-k}(j',j)$ at ${\bf q=k'-k}=(\pi,\pi)$. Hence a large absolute value of the matrix $M$ at {\bf Q} can lead to the superconducting transition. This value of $M$ stems from $V^s_{\bf Q}$ at small frequencies and, therefore, it has to correlate with the zero-frequency staggered spin susceptibility. At half-filling, this quantity diverges as $T$ tends to zero, indicating the transition to the long-range antiferromagnetic order. However, at finite doping, its value and thermal growth are limited due to the gap at {\bf Q} in the spin-excitation spectrum, with the gap value determined by the doping level \cite{Sherman03}. Experimentally, this fact manifests itself in the spin correlation length, which ceases to depend on the temperature at low $T$, being defined by the doping level \cite{Keimer}. Comparison of panels (a) and (b) in figure~\ref{Fig4} shows that there exists a moderate growth of $|M_{\bf Q}|$ with decreasing temperature. However, in accordance with figure~\ref{Fig2}, this growth is not enough for the eigenvalue in (\ref{Eliash}) to reach unity. It appears that {\it the necessary condition for the appearance of conductivity -- doping -- affects adversely on superconductivity}.

\begin{figure}[tbh]
\centerline{\resizebox{0.99\columnwidth}{!}{\includegraphics{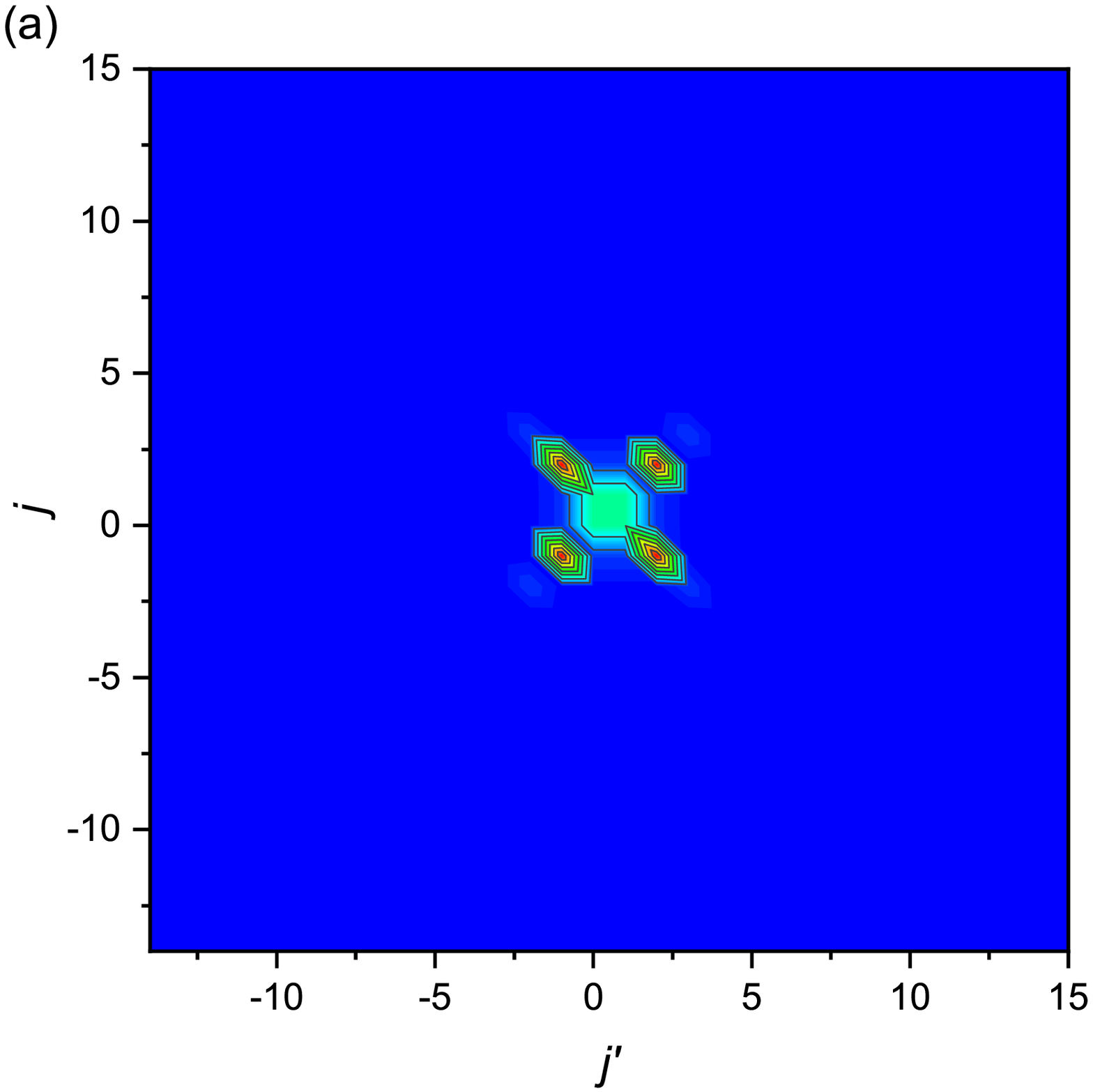}}}
\vspace{1ex}
\centerline{\resizebox{0.99\columnwidth}{!}{\includegraphics{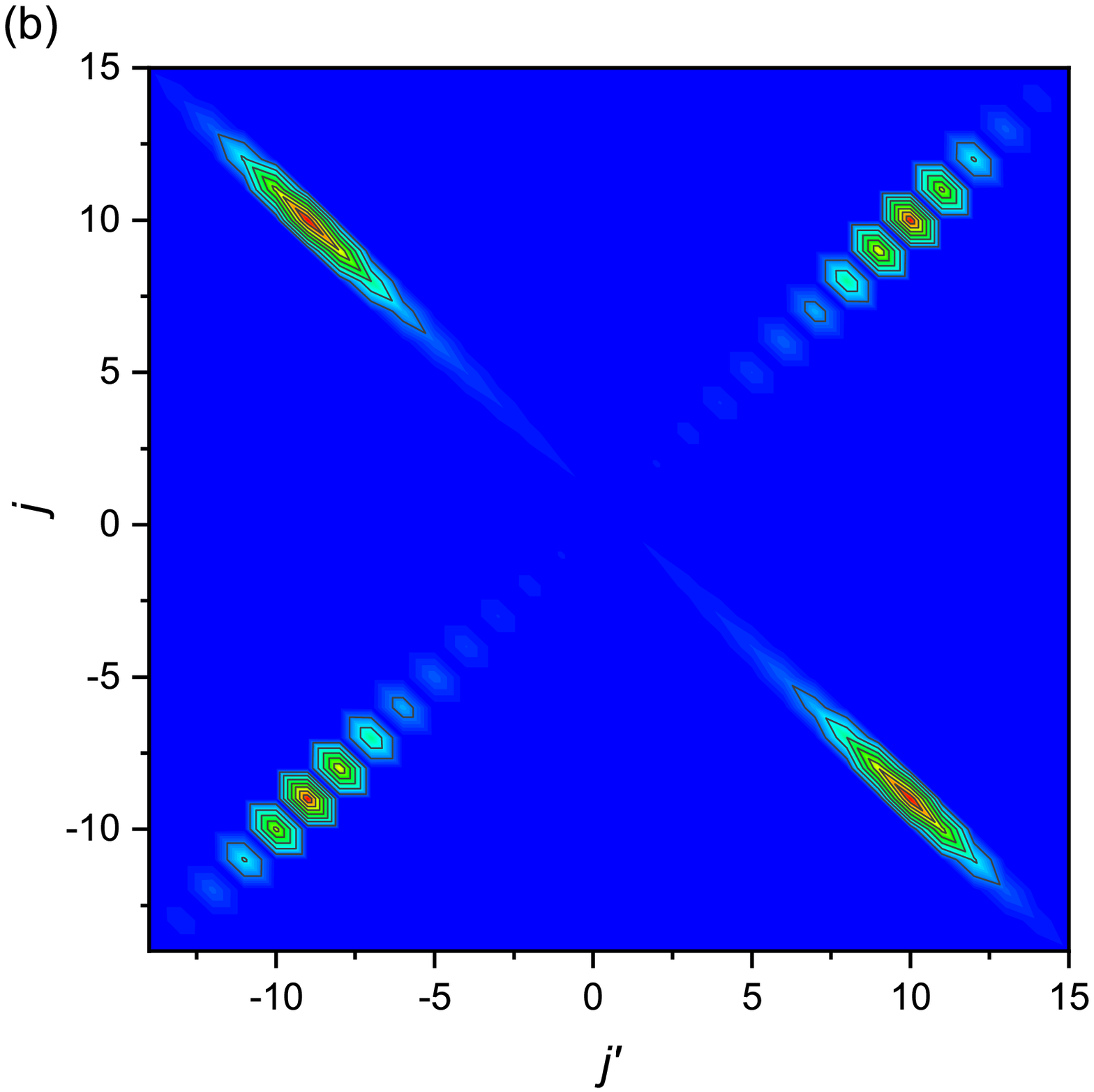}}}
\caption{The contour plot of the dependence ${\cal E}(j',j)$ (\protect\ref{contrib}) for $U=8t$, $t'=-0.3t$, $t''=0.2t$, $\bar{n}=0.92$, and the singlet $B_1$ pairing. Temperatures for panels (a) and (b) are $T\approx 0.13t$ and $0.02t$, respectively.} \label{Fig5}
\end{figure}

\begin{figure*}[t]
\centerline{\resizebox{0.9\columnwidth}{!}{\includegraphics{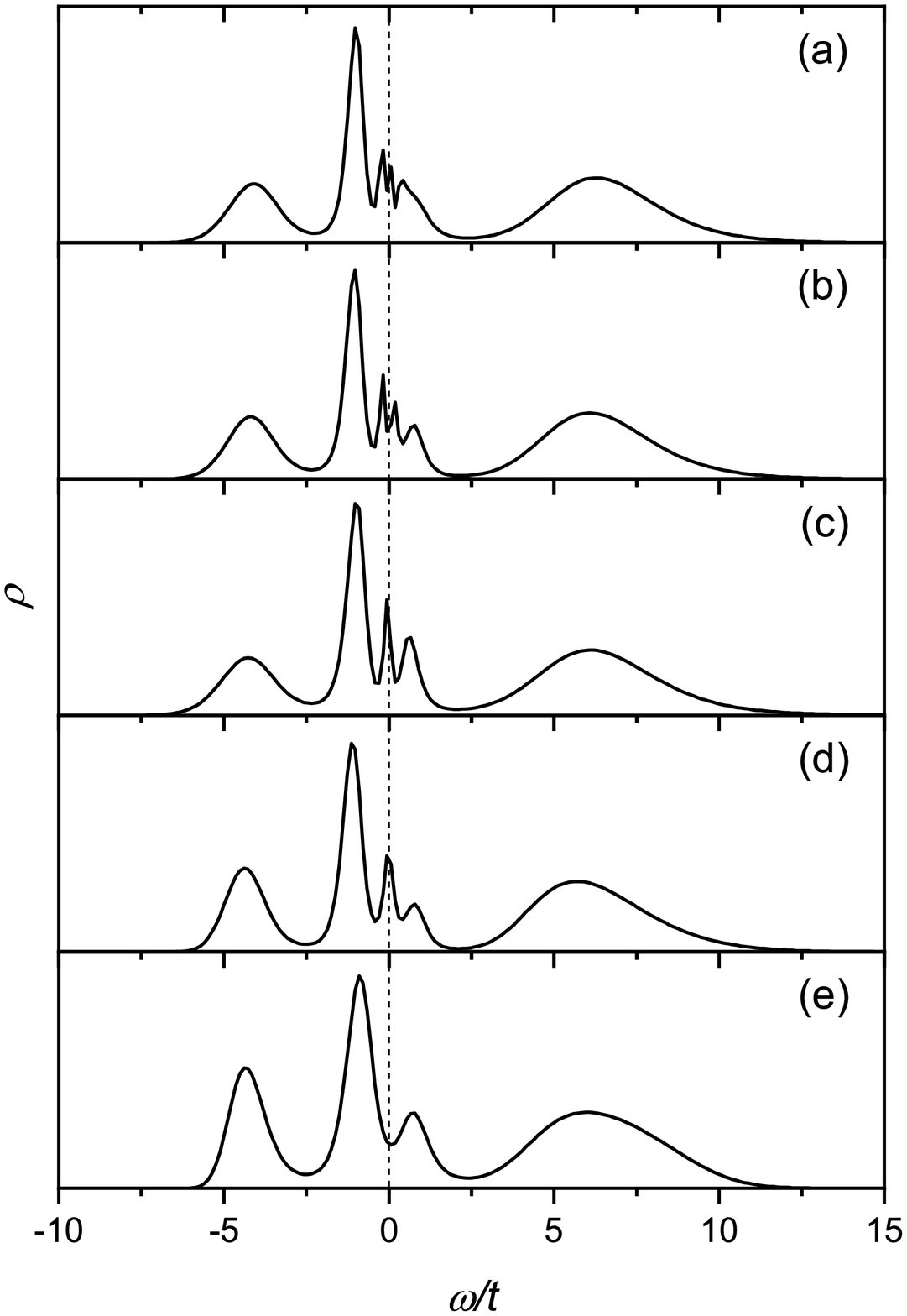}}\hspace{1em} \resizebox{0.92\columnwidth}{!}{\includegraphics{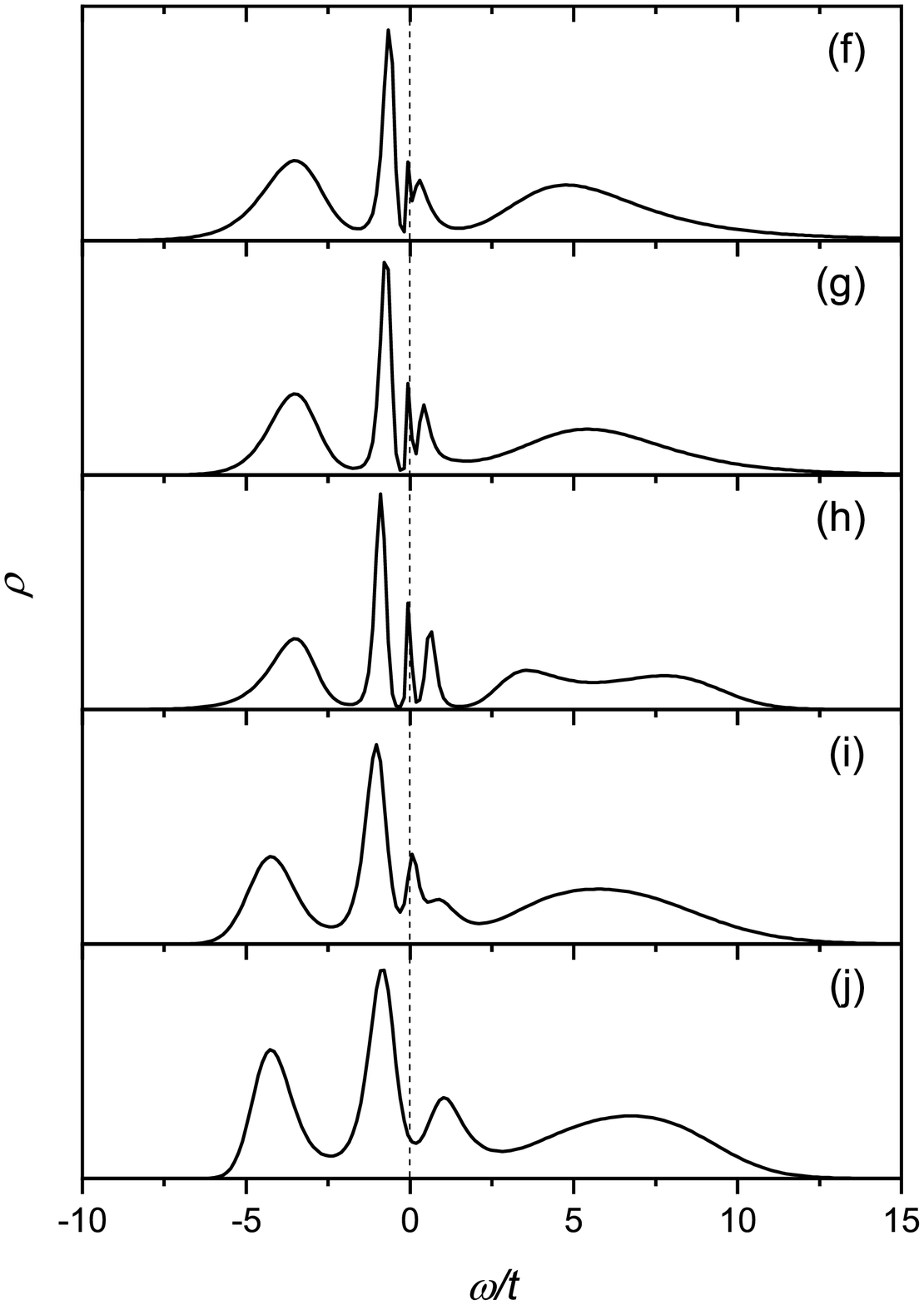}}}
\caption{Densities of states for the $t$-$U$ model with $U=8t$ (a-e) and the $t$-$t'$-$t''$-$U$ model with $U=8t$, $t'=-0.3t$, and $t''=0.2t$ (f-j). For both models $\bar{n}=0.92$. The temperature equals to $0.032t$ in panels (a) and (f), $0.051t$ in (b) and (g), $0.085t$ in (c) and (h), $0.13t$ in (d) and (i), and $0.32t$ in (e) and (j).} \label{Fig6}
\end{figure*}

As seen in figure~\ref{Fig2}(b), in spite of the mentioned moderate growth of $|M_{\bf Q}(0,0)|$ with decreasing temperature, the eigenvalue corresponding to the singlet $B_1$ ($d_{x^2-y^2}$) pairing decreases with $T$. Similar behavior is observed for some other symmetries also. To clarify this issue, let us consider the quantity
\begin{equation}\label{contrib}
{\cal E}(j',j)=\frac{\sum_{\bf k'k}\varphi^*_{\bf k'}(j')M_{\bf k'-k}(j',j)\varphi_{\bf k}(j)}{\sum_{{\bf k}j}\varphi^*_{\bf k}(j)\varphi_{\bf k}(j)},
\end{equation}
which shows the contribution of different frequency regions to an eigenvalue. The contour plot of this quantity is shown in figure~\ref{Fig5} for two values of $T$. The value of ${\cal E}$ grows from blue to red regions. As seen from the figure, frequency regions making the main contribution to $E$ are essentially changed with temperature. For larger $T$, they are located at low frequencies, while for smaller temperatures the higher frequencies contribute the most. As follows from figure~\ref{Fig3}, the vertex $V^+$ is larger in the former region and smaller in the latter, which explains the decrease of the eigenvalue with temperature. The thermal changes in ${\cal E}$ may point to a modification in retardation effects occurring in the system with the temperature variation.

Densities of states,
\begin{equation}\label{dos}
\rho(\omega)=-\frac{1}{N\pi}\sum_{\bf k}\int_{-\infty}^{\infty}{\rm Im}\,G({\bf k},\omega)\,{\rm d}\omega,
\end{equation}
calculated from electron Green's functions obtained in this work are shown in figure~\ref{Fig6}. In (\ref{dos}), the electron Green's function $G({\bf k},\omega)$ on the real frequency axis was derived from the self-consistently calculated Matsubara function using the maximum entropy method \cite{Press,Jarrell,Habershon}. A noteworthy feature of these spectra is the sharp peak at the Fermi level, which arises at a low enough temperature. It is connected with flat bands formed by bound states of electrons and spin excitations \cite{Sherman19a,Sherman19b}. By analogy with similar excitations of the $t$-$J$ model \cite{Schmitt,Ramsak,Sherman94} they were named the spin-polaron states. As seen from figure~\ref{Fig6}, they persist up to the lowest considered temperatures. Analogous peaks were observed in densities of states of the three-band (Emery) and Hubbard-Kanamori models \cite{Sherman20a,Sherman20b}. Some peculiarities in photoemission spectra of $n$- and $p$-type cuprates can supposedly be identified with these peaks \cite{Armitage01,Armitage02,Damascelli,Matsui05,Matsui}. Such Fermi-level peaks can significantly increase the transition temperature of the phonon-mediated superconductivity \cite{Abrikosov}.

\section{Conclusion}
In this work, the possibility of the superconducting transition was investigated in the two-dimensional fermionic Hubbard model. Two types of kinetic-energy terms were considered. In one of them, only the nearest-neighbor hopping constant $t$ is nonzero; the other contains nonzero hopping integrals between second and third neighbors also. Their values were determined in band-structure calculations. We considered the case of strong electron correlations with the Coulomb on-site repulsion $U=8t$. The calculations were carried out using the strong coupling diagram technique with the self-consistently obtained electron Green's function. Spectral and magnetic properties of the model calculated in this approach previously were shown to be in good agreement with the results of numeric experiments and experiments with ultracold atoms in optical lattices. To avoid the influence of the charge separation and size effects calculations were performed in an infinite crystal for the electron concentration $\bar{n}=0.92$, in the part of the phase diagram, which is free from inhomogeneities. In solving the Eliashberg equation, the singlet and triplet pairing and eigenfunctions of all one-dimensional representations of the lattice point group $D_4$ were considered. In the range of temperatures $0.02t\lesssim T\lesssim 0.32t$, the eigenvalues of the Eliashberg equation appeared to be less than unity for both models and all considered symmetries. It points to the absence of the superconducting transition in the Hubbard model in the strong-coupling regime. Notice that for the superexchange constant $J=4t^2/U=0.1$~eV, as in cuprates, and the considered value of $U$, the lower boundary of the used temperature range is close to the superconducting transition temperature in La$_{2-x}$Ba$_x$CuO$_4$. In the case of primary interest -- the singlet $d_{x^2-y^2}$ pairing -- the small eigenvalue of the Eliashberg equation is connected with the modest value of the low-frequency spin vertex near the antiferromagnetic momentum and its moderate growth with decreasing temperature at finite doping. This thermal behavior of the spin vertex reveals itself also in the saturation of zero-frequency staggered susceptibility and in the spin correlation length, which, at low $T$, ceases to depend on temperature and is determined by doping only.


\section*{References}
\begin{thebibliography}{99}
\bibitem{Anderson}Anderson P W 1987 {\it Science} {\bf 235} 1196
\bibitem{Zhang}Zhang F C and Rice T M 1988 {\it Phys. Rev.} B {\bf 37} 3759
\bibitem{Hirsch}Hirsch J E 1985 {\it Phys. Rev.} B {\bf 31} 4403
\bibitem{Scalapino}Scalapino D J 2012 {\it Rev. Mod. Phys.} {\bf 84} 1383
\bibitem{Otsuki}Otsuki J, Haffermann H and Lichtenstein A I 2014 {\it Phys. Rev.} B {\bf 90} 235132
\bibitem{Aichhorn}Aichhorn M, Arrigoni E, Potthoff M and Hanke W 2006 {\it Phys. Rev.} B {\bf 74} 235117
\bibitem{Macridin}Macridin A, Jarrel M and Maier T 2006 {\it Phys. Rev.} B {\bf 74} 085104
\bibitem{White}White S R and Scalapino D J 1998 {\it Phys. Rev. Lett.} {\bf 80} 1272
\bibitem{Sherman18}Sherman A 2018 {\it J. Phys.: Condens. Matter} {\bf 30} 195601
\bibitem{Sherman19a}Sherman A 2019 {\it Eur. Phys. J.} B {\bf 92} 55
\bibitem{Sherman20}Sherman A 2020 {\it Phys. Scr.} {\bf 95} 015806; arXiv:2010.00218
\bibitem{Bill}Bill A, Hizhnyakov V, Kremer R K, Seibold G, Shelkan A and Sherman A 2020 {\it Condensed Matter} {\bf 5} 65
\bibitem{Senechal}S\'en\'echal D, Lavertu P.-L., Marois M.-A. and Tremblay A.-M. S. 2005 {\it Phys. Rev. Lett.} {\bf 94} 156404
\bibitem{Capone}Capone M and Kotliar G 2006 {\it Phys. Rev.} B {\bf 74} 054513
\bibitem{Aimi}Aimi T and Imada M 2007 {\it J. Phys. Soc. Jpn} {\bf 76} 113708
\bibitem{Maier}Maier T A, Jarrel M S and Scalapino D J 2006 {\it Phys. Rev. Lett.} {\bf 96} 047005
\bibitem{Kitatani}Kitatani M, Sch\"afer T, Aoki H and Held K 2019 {\it Phys. Rev.} B {\bf 99} 041115(R)
\bibitem{Qin}Qin M, Chung C.-M., Shi H, Vitali E, Hubig C, Schollw\"ock U, White S R and Zhang S 2020 {\it Phys. Rev.} X {\bf 10} 031016
\bibitem{Tranquada}Tranquada J M, Sternlieb B J, Axe J D, Nakamura Y and Uchida S 1995 {\it Nature} {\bf 375} 561
\bibitem{Forgan}Forgan E M, Blackburn E, Holmes A T, Briffa A K R, Chang J, Bouchenoire L, Brown S D, Liang R, Bonn D, Hardy W N, Christensen N B, v. Zimmermann M, H\"ucker M and Hayden S M 2015 {\it Nat. Commun.} {\bf 6} 10064
\bibitem{Eliashberg}Eliashberg G M 1960 {\it Soviet Phys. JETP} {\bf 11} 696
\bibitem{Vladimir}Vladimir M I and Moskalenko V A  1990 {\it Theor.\ Math.\ Phys.} {\bf 82} 301
\bibitem{Metzner}Metzner W 1991 {\it Phys.\ Rev.} B {\bf 43} 8549
\bibitem{Pairault}Pairault S, S\'en\'echal D and Tremblay A-M S 2000 {\it Eur.\ Phys.\ J.} B {\bf 16} 85
\bibitem{Sherman16}Sherman A 2016 {\it Eur. Phys. J.} B {\bf 89} 91
\bibitem{Zaitsev}Zaitsev R O 1976 {\it Sov.\ Phys.\ JETP} {\bf 43} 574
\bibitem{Izyumov}Izyumov Yu A and Skryabin Yu N 1988 {\it Statistical Mechanics of Magnetically Ordered Systems} (New York: Consultants Bureau)
\bibitem{Sherman19b}Sherman A 2019 {\it Phys.\ Scr.} {\bf 94} 055802
\bibitem{Andersen}Andersen O K, Liechtenstein A I, Jepsen O and Paulsen F 1995 {\it J. Phys. Chem. Solids} {\bf 56} 1573
\bibitem{Keimer}Keimer B, Belk N, Birgeneau R J, Cassanho A, Chen C Y, Greven M, Kastner M, Aharony A, Endoh Y, Erwin R and Shirane G 1992 {\it Phys. Rev.} B {\bf 46} 14034
\bibitem{Hubbard63}Hubbard J 1963 {\it Proc.\ R.\ Soc.\ Lond.} A {\bf 276} 238
\bibitem{Hubbard64}Hubbard J 1964 {\it Proc.\ R.\ Soc.\ Lond.} A {\bf 277} 237
\bibitem{Kubo}Kubo R 1962 {\it J. Phys. Soc. Jpn.} {\bf 17} 1100
\bibitem{Bir}Bir G L and Pikus G E 1974 {\it Symmetry and strain-induced effects in semiconductors} (New York: Wiley)
\bibitem{Mises}v. Mises R and Pollaczek-Geiringer H 1929 {\it Zeit\-schrift f\"ur Angewandte Mathematik und Mechanik} {\bf 9} 152
\bibitem{Sherman03}Sherman A and Schreiber M 2003 {\it Eur. Phys. J.} B {\bf 32} 203
\bibitem{Press}Press W H, Teukolsky S A, Vetterling W T and Flannery B P 1995 {\it Numerical Recipes in Fortran} (Cambridge: Cambridge University Press) chapter 18
\bibitem{Jarrell}Jarrell M and Gubernatis J E 1996 {\it Phys.\ Rept.} {\bf 269} 133
\bibitem{Habershon}Habershon S, Braams B J and Manolopoulos D E 2007 {\it J.\ Chem.\ Phys.} {\bf 127} 174108
\bibitem{Schmitt}Schmitt-Rink S, Varma C M and Ruckenstein A E 1988 {\it Phys. Rev. Lett.} {\bf 60} 2793
\bibitem{Ramsak}Ramšak A and Horsch P 1993 {\it Phys. Rev.} B {\bf 48} 10559
\bibitem{Sherman94}Sherman A and Schreiber M 1994 {\it Phys. Rev.} B {\bf 50} 12887
\bibitem{Sherman20a}Sherman A 2020 {\it Eur. Phys. J.} B {\bf 93} 168
\bibitem{Sherman20b}Sherman A 2020 {\it Phys. Scr.} {\bf 95} 095804
\bibitem{Armitage01}Armitage N P, Lu D H, Kim C, Damascelli A, Shen K M, Ronning F, Feng D L, Bogdanov P, Shen Z-X, Onose Y, Taguchi Y, Tokura Y, Mang P K, Kaneko N and Greven M 2001 {\it Phys. Rev. Lett.} {\bf 87} 147003 (2001).
\bibitem{Armitage02}Armitage N P, Ronning F, Lu D H, Kim C, Damascelli A, Shen K M, Feng D L, Eisaki H, Shen Z-X, Mang P K, Kaneko N, Greven M, Onose Y, Taguchi Y and Tokura Y 2002 {\it Phys. Rev. Lett.} {\bf 88} 257001
\bibitem{Damascelli}Damascelli A, Hussain Z and Shen Z-X 2003 {\it Rev.\ Mod.\ Phys.} {\bf 75} 473
\bibitem{Matsui05}Matsui H, Terashima K, Sato T, Takahashi T, Wang S-C, Yang H-B, Ding H, Uefuji T and Yamada K 2005 {\it Phys. Rev. Lett.} {\bf 94} 047005
\bibitem{Matsui}Matsui H, Takahashi T, Sato T, Terashima K, Ding H, Uefuji T and Yamada K 2007 {\it Phys.\ Rev.} B {\bf 75} 224514
\bibitem{Abrikosov} Abrikosov A A, Gor’kov L P and Dzyaloshinskii I E 1965 {\it Methods of Quantum Field Theory in Statistical Physics} (New York: Pergamon Press)
\end{thebibliography}
\end{document}